# Unravelling the origin of piezo/ferro-electric properties of metal-organic frameworks (MOFs) nanocrystals

*Yao Sun,* [†, ‖] *Junfeng Gao,* [‡, ‖] *Yuan Cheng,* [‡, *] *Yong-Wei Zhang,* [‡] *and Kaiyang Zeng* [†, *]

[†]Department of Mechanical Engineering, National University of Singapore, 9 Engineering Drive 1, 117576, Singapore

[‡]Institute of High Performance Computing, Agency for Science Technology and Research, 1 Fusionopolis Way, 138632, Singapore

*Abstract*

Metal-organic framework (MOF) UiO-66 nanocrystals were previously believed to be piezo/ferro-electrically inactive because of their centrosymmetric lattice symmetries (*Fm-3m (225)*) revealed by Powder X-ray diffraction. However, *via* delicate dual AC resonance tracking piezoresponse force microscopy and piezoresponse force spectroscopy characterizations, our nanoscale probing for the first time demonstrate that UiO-66 nanocrystals show piezo/ferro-electric response. Our compelling experimental and theoretically analyses disclose that the structure of UiO-66 should not be the highly centrosymmetric *Fm-3m (225)* but a reduced symmetry form instead. UiO-66(Hf)-type MOFs possess stronger piezoresponse and better ferroelectric switching behaviours than their counterparts UiO-66 (Zr)-type MOFs. Our study not only enriches the structural understanding of UiO-66 MOF, but also suggests possible modification of electronic property of the MOFs by judicious selection of metal ions and functional ligands.

*Introduction*

Piezoelectric depicts the ability of materials that can convert mechanical energy to electrical energy and *vice versa*.[1] In general, piezoelectric materials have a non-centrosymmetric crystalline structure, which gives rise to the dielectric polarization when subject to stress. Ferroelectric arises when piezoelectric materials are capable of developing a spontaneous and reversible dielectric polarization. These materials serve as a fundamental element of electronic devices, micro-mechanics and memories,[2, 3, 4, 5, 6, 7, 8] such as actuators, MEMS, electro-optics, cooling and electron emitters and FeRAM (Ferroelectric random access memory), *etc*. In addition, the dependence of piezo/ferro-electric properties on the symmetry group provides an effective criterion to distinguish the crystalline structure with or without centrosymmetric.

The UiO-66, the prototype of zirconium-based metal−organic frameworks (MOFs), is featured by unprecedented stability, accessible fabrication, high surface area (up to 7000 $m^2/g$)[9], tunable pore aperture (up to 98 Å)[10] and crystal density (as low as 0.13 $g/cm^3$)[11]. It exhibits brilliant prospects in gas storage and separation,[12, 13] catalysis,[14, 15] chemical sensing,[16] drug delivery,[17] and photochemical application,[18, 19, 20] and more. By Powder X-Ray Diffraction (PXRD) measurement, the UiO-66 was thought to be highly centrosymmetric *[Fm-3m (225)]*.[21, 22] However, the precision of PXRD may be not enough to identify the small motifs and slight structure distortions in UiO-66. Besides, although the perfect UiO-66 crystal has a 12-connected framework structure, it was proposed a number of missing-linker were always present as evidenced by thermogravimetric analysis.[23, 24] Recently, De Vos *et al.* have theoretically indicated that the properties of UiO-66 can be altered remarkably by the missing-linker and structural distortion.[25] Combing Extended X-Ray Absorption Fine Structure (EXAFS) analysis and *ab initio* calculations, the low symmetric structure *[F-43m (216)]* was proposed for the structures of UiO-66.[23]

However, the crystal structure of UiO-66 is still commonly used in the following theoretical *[F-43 (216)]*[23, 26, 27] and experimental works *[Fm-3m (225)]*[22, 23, 28]. Further straightforward and convictive experimental observations are necessary to identify the real crystal structure of UiO-66. Obviously, highly centrosymmetric *Fm-3m (225)* must be piezo/ferro-electrically inactive, while lower symmetric structures such as *F-43m (216)* may be piezo/ferro-electrically active. Therefore, piezo/ferro-electric response observation may be a cogent external method to verify the crystalline structure.

Beyond the fundamental understanding of crystal structure of UiO-66, the inorganic-organic hybrid ferroelectrics have emerged as a new research frontier in materials science recently. Compared to conventional ferroelectrics, inorganic-organic hybrid ferroelectrics combine the advantages of both organic linkers, such as straightforward synthesis and easily tailored molecular structure, and inorganic bricks, and thus have favourable chemical, thermal and mechanical stabilities.[29] it is noted that by tuning the reaction conditions, one can easily design the desired coordination frameworks and obtain a large number of coordination polymers,[30] paving the way to develop new high-performance functional materials in the near future. For example, hybrid perovskite methylammonium lead trihalide ($MAPbX_3$) thin films provide an alternative but exhilarating solution for high-performance ferroelectric solar cells beyond inorganic ferroelectric oxides.[31] Coordination polymers with dynamic micropores coupled with guest occlusion are found to be promising materials for applications in sensors and actuators. However, current research on UiO-66 ferroelectrics is still largely unexplored.

The Dual AC Resonance Tracking Piezoresponse Force Microscopy (DART-PFM)[32] and Piezoresponse Force Spectroscopy (PFS)[33] are the emerging techniques that can be employed to measure the piezo/ferro- electric property of materials at the nanoscale. PFM as well as its spectroscopic mode have been applied to characterize a wide range of materials, including thin films,[34, 35] nanowires,[36] nanoparticles[37] and other confined systems,[38, 39] *etc*. To the best of our knowledge, PFM and PFS have rarely been applied to characterize the piezo/ferro-electric properties of MOFs. Our recent work has adopted DART-PFM and PFS to successfully reveal that the nanoscale piezo/ferro-electric active behavior of NUS-6 MOF is due to a charged asymmetric crystal structure caused by the consistence of missing ligands and clusters.[40] Apart from experimental works, calculating Born effective charge tensor is one of the most widely acknowledged methods for ferroelectric phases exploration of crystalline materials.[41, 42, 43] At present stage, the first-principles strategy on MOFs ferroelectricity remains quite limited because of large demand of computing capacity, especially for the MOFs with large atomic numbers.[44, 45] In this study, we explore the piezo/ferro-electric behavior of UiO-66-type MOFs using DART-PFM and PFS incorporated with first-principles calculations. The unexpected piezo/ferro-electricity is observed in both UiO-66(Hf/Zr) MOFs, verifying the lower crystalline symmetry. Besides, both experimental observation and first-principles calculations indicate that the UiO-66(Hf)-type MOFs exhibit stronger piezo/ferro-electric responses than those of the UiO-66(Zr)-type MOFs. These results imply that a change of the inorganic metal modes and organic linkers can effectively tune piezo/ferro-electric responses of the UiO-66 MOFs, which may lead to versatile piezo/ferro-electrics.

*Results*

**DART-PFM measurements on UiO-66-type nanocrystals.**

For DART-PFM measurements, a driving voltage of 5 V is applied on the conductive probe (240AC-PP, OPUS, CA, USA). The DART-PFM images of four UiO-66(Hf)-type and four UiO-66(Zr)-type MOFs (Supplementary Fig. 1) nanocrystals are shown in Figs. 1 and 2, respectively.

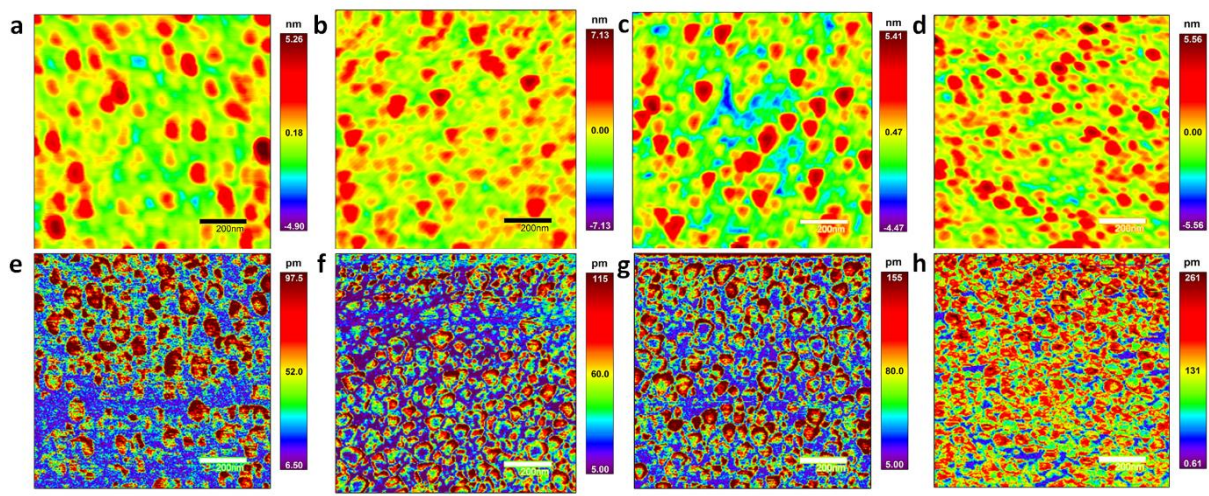

**Figure 1.** DART-PFM images of UiO-66(Hf)-type nanocrystals. **a-d** Topography images and **e-h** DART-PFM amplitude images. **a, e** UiO-66(Hf), **b, f** UiO-66(Hf)-NH$_2$, **c, g** UiO-66(Hf)-(OH)$_2$, and **d, h** UiO-66(Hf)-(COOH)$_2$.

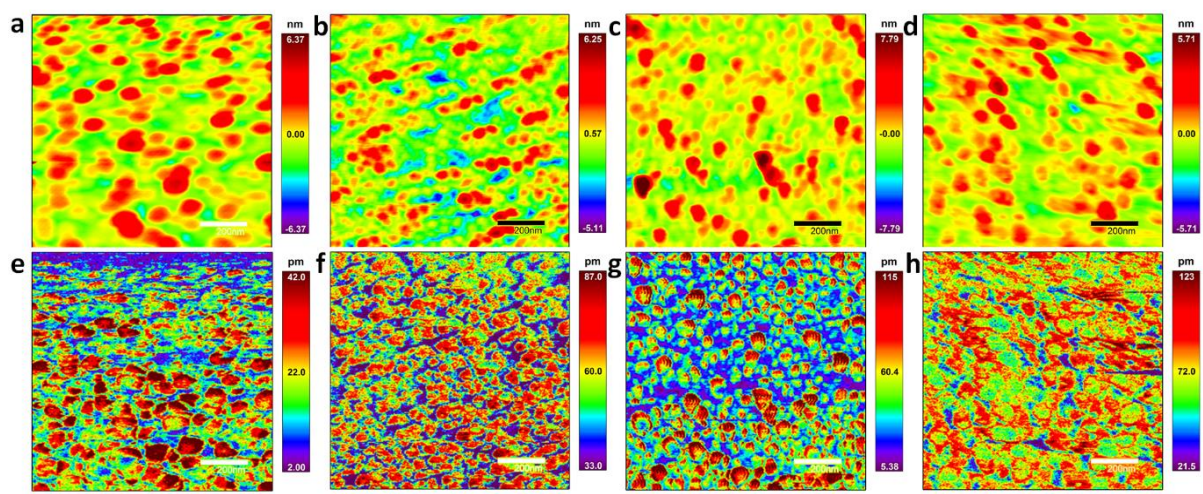

**Figure 2.** DART-PFM images of UiO-66(Zr)-type nanocrystals. **a-d** Topography images and **e-h** DART-PFM amplitude images. **a, e** UiO-66(Zr), **b, f** UiO-66(Zr)-NH$_2$, **c, g** UiO-66(Zr)-(OH)$_2$, and **d, h** UiO-66(Zr)-(COOH)$_2$.

From the topography images (Figs. 1a-d and Figs. 2a-d), it can be seen that the sizes of the UiO-66(Hf/Zr) MOF nanocrystals are approximately 100 ~ 200 nm with mild variations. Besides, from the PFM amplitude images (Figs. 1e-h and Figs. 2e-h), it is conspicuous that both UiO-66(Hf)- and UiO-66(Zr)-type nanocrystals exhibit piezoresponse with the amplitude of several $10^2$ picometer (pm) under the applied voltages. From those DART-PFM amplitude images (Figs. 1e-h and Figs. 2e-h), the non-uniform piezoresponse distribution in each nanocrystal can also be clearly seen, suggesting possible structural anisotropy within the UiO-66(Hf/Zr)-type MOFs nanocrystals. The PFM amplitudes of UiO-66(Hf)-type nanocrystals are approximately in the ranges of 6.5-97.5 (UiO-66(Hf)), 5-115 (UiO-66(Hf)-NH$_2$), 5-155 (UiO-66(Hf)-(OH)) and 0.61-261 pm (UiO-66(Hf)-(COOH)$_2$) (Figs. 1e-h), respectively. While the ranges of PFM amplitudes for Zr-based UiO-66 are approximately 2-42 (UiO-66(Zr)), 33-87 (UiO-66(Zr)-NH$_2$), 5.38-115 (UiO-66(Zr)-(OH)$_2$) and 21.5-123 pm (UiO-66(Zr)-(COOH)$_2$), respectively (Figs. 2e-h). Obviously, the UiO-66(Hf)-type nanocrystals show larger piezoresponse with the maximum PFM amplitude of 97.5-261 pm, compared to the maximum amplitude (42-123 pm) in the UiO-66-Zr-type nanocrystals under the same driving voltages. It is worthy to notice that the UiO-66(Hf/Zr)-type MOFs with functional groups -NH$_2$, -OH, and -COOH tend to show larger piezoresponse than that of the corresponding pristine MOFs. This may be due to the fact that the permanently charged functional groups, such as -COOH, -NH$_2$ and phenolic -OH, can increase the polarity of the overall structures.[46]

**PFS measurements on UiO-66-type nanocrystals.**

The PFS driving voltage applied is comprised of an ac voltage (5V: PFM driving amplitude) coupled with a specific dc voltage. The recorded four cycles butterfly-shaped amplitude loops and the corresponding phase loops of UiO-66-type MOFs nanocrystals at low ($V_{dc}=0$) state are shown in Supplementary Fig. 2.

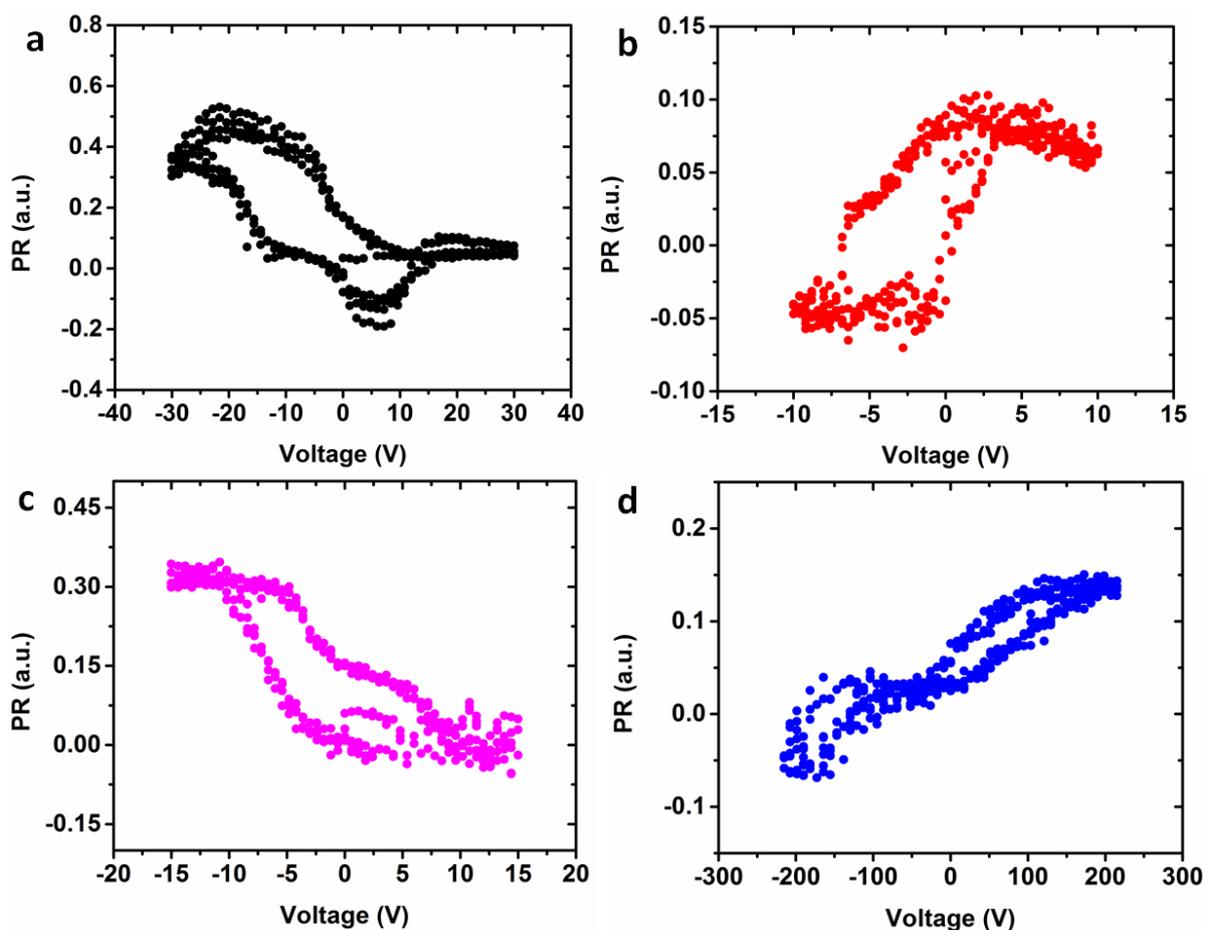

**Figure 3.** Ferroelectric hysteresis loops (PR loops) in PFS measurements. **a** UiO-66(Hf), **b** UiO-66(Hf)-NH$_2$, **c** UiO-66(Hf)-(OH)$_2$, and **d** UiO-66(Zr)-(OH)$_2$. All the PR loops are non-symmetric and shifted along the voltage axis.

All the UiO-66(Hf)-type MOF nanocrystals demonstrate delicate butterfly-shaped amplitude loops (Supplementary Figs. 2a, c, e and g) and corresponding phase loops (Supplementary Figs. 2b, d, f and h). Whereas for UiO-66(Zr)-type MOF nanocrystals, only UiO-66(Zr)-NH$_2$ and UiO-66(Zr)-(OH)$_2$ show the butterfly-shaped amplitude loops (Supplementary Figs. 2i and k) and corresponding phase loops (Supplementary Figs. 2j and l). After calculation of PR loops from amplitude butterfly (A) and phase loops (ϕ) [PR loops: PR= A×cos(ϕ)] (Supplementary Fig. 2), it is noticed that UiO-66(Hf)-(COOH)$_2$ and UiO-66(Zr)-NH$_2$ have no ferroelectric polarization switching behaviors since their ferroelectric hysteresis loops (PR loops) only remain at the positive level (PR > 0) (Supplementary Fig. 3). For UiO-66(Zr) and UiO-66(Zr)-(COOH)$_2$ nanocrystals, neither piezoelectric butterfly nor phase loops can be detected.

In addition, Fig. 3 shows the PR loops obtained from UiO-66(Hf), UiO-66(Hf)-NH$_2$, UiO-66(Hf)-(OH)$_2$ and UiO-66(Zr)-(OH)$_2$ nanocrystals. It can be seen that the PR loops are non-symmetric and shifted along the voltage axis, indicating that internal bias exist due to the aligned dipoles.[47] In the structures of UiO-66(Hf/Zr)-type MOFs, the metal (Hf/Zr)-Oxygen coordination system is supposed to act as active dipoles.[51] Indeed, it was found that non-covalent bonds have the polar nature because of the asymmetry of electron density configurations when two or more molecules interact with each other.[48] For UiO-66(Hf)-type MOFs, UiO-66(Hf), UiO-66(Hf)-NH$_2$, UiO-66(Hf)-(OH)$_2$ all show exquisite PR loops under the driving voltage of 30V (Fig. 3a), 10V (Fig. 3b) and 15V (Fig. 3c). However, for UiO-66(Zr)-type MOFs, only UiO-66(Zr)-(OH)$_2$ shows barely satisfactory PR loop (Fig. 3d) under a very large driving voltage (216V).

**Polarization calculations of UiO-66.**

In the calculation, primitive cell of UiO-66(Hf/Zr) including 114 atoms are adopted instead of the unit cell (456 atoms) in order to simplify the calculation processes. The unit cell of UiO-66(Zr) is also calculated, and the result is consistent with that of primitive cell. The primitive cell volumes ($\Omega$) of UiO-66(Hf) and UiO-66(Zr) turn to be 2273.04 Å$^3$ and 2306.89 Å$^3$ after structural optimizations. Figure 4 shows the primitive cells (Figs. 4a and b) and optimized unit cell structures (Figs. 4d and e) of UiO-66(Zr) and UiO-66(Hf), respectively. The corresponding secondary building units (SBUs) depict the intrinsic difference of metal-oxygen bonding (Fig. 4c), demonstrating the shorter bond length of the Hf-O than that of the Zr-O. After structure relaxation, the lattice symmetries of both UiO-

66(Zr) and UiO-66(Hf) are the *F-43m (216)*, in agreement with previous proposed structures.[23]

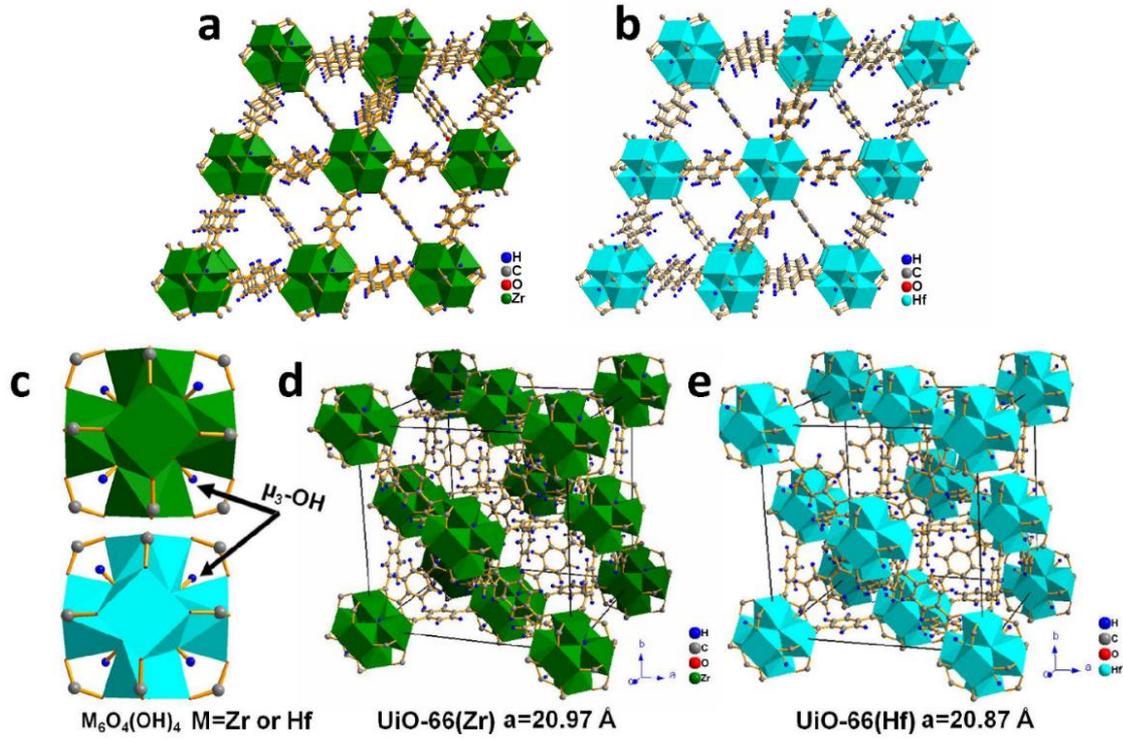

**Figure 4.** The structural descriptions. **a, c, d** UiO-66(Zr) and **b, c, e** UiO-66(Hf). **a-b** Primitive cell, **c** Secondary building unit, **d-e** Optimized unit cell.

The calculated Born effective charge $Z^{*}$[42, 43, 49] tensors for both UiO-66(Hf) and UiO-66(Zr) are in forms of asymmetric full matrix due to the low symmetry of optimized structures of face-centred cubic structure *(FCC) F-43m (216)*.[26] For instance,

$$Z^{*}(Hf)_{109} = \begin{pmatrix} 4.26371 & 0.00000 & 0.00000 \\ -0.00001 & 5.21418 & -0.36317 \\ -0.00004 & -0.36314 & 5.21406 \end{pmatrix}, \quad Z^{*}(Zr)_{109} = \begin{pmatrix} 4.34547 & -0.00000 & -0.00000 \\ -0.00086 & 5.12270 & -0.38090 \\ -0.00115 & -0.38098 & 5.12308 \end{pmatrix}.$$

Herein, the number 109 indicates the atom index of Hf or Zr atom. Table 1 shows the results of Born effective charge $Z^*$ and polarization distance $\Delta u$ of Hf and Zr atoms in the z direction. The Born effective charge $Z^*$ tensors for each atom in UiO-66(Hf) and UiO-66(Zr)'s primitive cells are available in the Supplementary Tables 1 and 2. While the Born effective charge $Z^*$ and polarization distance of other comprising atoms (C, H, O) in UiO-66(Hf) and UiO-66(Zr) in the z direction are listed in Supplementary Tables 3 and 4. It is clear that the Born effective charge $Z^*$ of Hf is slightly larger than that of Zr. Furthermore, particularly, the polarization distance $\Delta u$ of UiO-66(Hf) is almost twice than that of UiO-66(Zr).

**Table 1.** Born effective charge $Z^*$ and polarization distance $\Delta u$ data of Hf/Zr in UiO-66.

|    | Atom $i$ | Polarization distance $\Delta u$ | Born effective charge $Z^*$ | ionic |
|----|------|------------------|--------------------------|----|
| **Hf** | 109 | 0, 0, -0.431 | 0.000, -0.363, 5.214 | 12 |
|    | 110 | 0, 0, -0.431 | 0.000, 0.363, 5.214 | 12 |
|    | 111 | 0, 0, -0.431 | -0.363, -0.000, 5.214 | 12 |
|    | 112 | 0, 0, -0.431 | 0.363, 0.000, 5.214 | 12 |
|    | 113 | 0, 0, -0.431 | 0.000, 0.000, 4.264 | 12 |
|    | 114 | 0, 0, -0.431 | 0.000, 0.000, 4.264 | 12 |
| **Zr** | 109 | 0, 0, -0.268 | -0.001, -0.381, 5.123 | 12 |
|    | 110 | 0, 0, -0.268 | 0.001, 0.381, 5.123 | 12 |
|    | 111 | 0, 0, -0.268 | -0.381, -0.001, 5.123 | 12 |
|    | 112 | 0, 0, -0.268 | 0.381, 0.001, 5.123 | 12 |
|    | 113 | 0, 0, -0.268 | 0.000, 0.000, 4.347 | 12 |
|    | 114 | 0, 0, -0.268 | -0.000, 0.000, 4.341 | 12 |

The calculation results show that the Born effective charges $Z^*$ of the six Hf/Zr atoms have different values in the z direction, demonstrating the presence of atomic geometry distortions.[42] The calculated polarization changes $(\Delta P)$[42] of the six Hf atoms ($i$=109, 110, 111, 112, 113, 114) in the primitive cell structure of the UiO-66(Hf) are specifically -0.190, -0.190, -0.190, -0.190, -0.155, and -0.155 eÅ, and the calculated polarization changes $(\Delta P)$ are -0.114, -0.114, -0.114, -0.114, -0.097, and -0.097 eÅ for the six Zr atoms ($i$=109, 110, 111, 112, 113, 114) in the UiO-66(Zr), respectively.

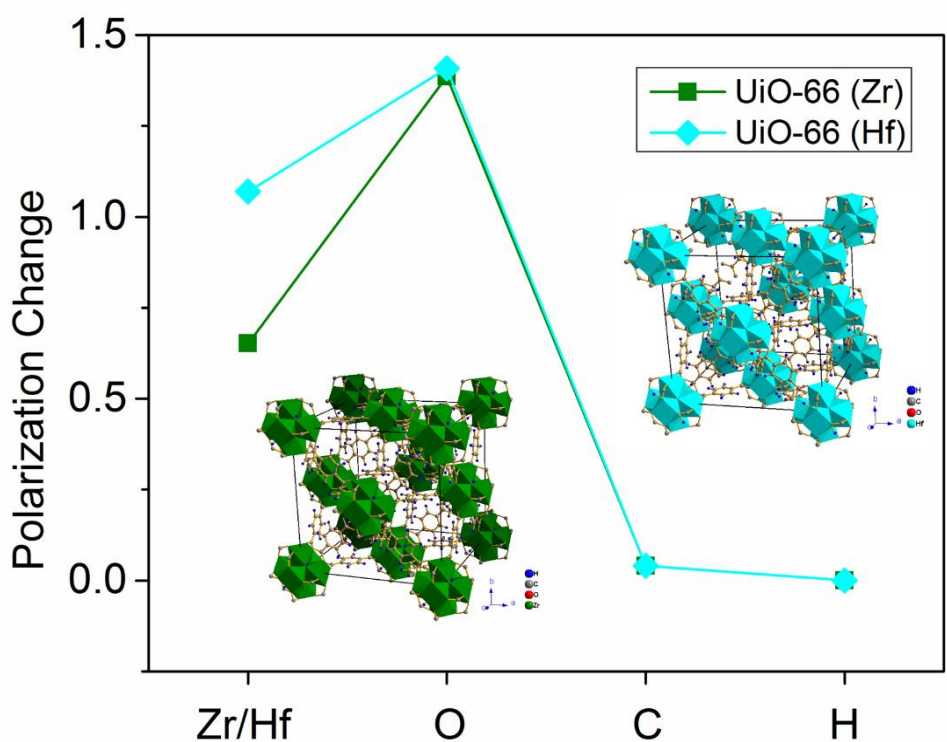

**Figure 5.** Polarization changes $\Delta P$ of the four atoms of UiO-66(Zr) and UiO-66(Hf).

Figure 5 shows the total polarization changes $(\Delta P)$ of the four comprising atoms for UiO-66(Hf) and UiO-66(Zr). The absolute value of total polarization change $|\Delta P|$ of Hf is found to be |1.070| eÅ, which is |0.419| eÅ larger than that of Zr (|0.651| eÅ). Here, the sign of polarization indicates the polarization direction. Similarly, the total polarization change $|\Delta P|$ of O atoms in UiO-66(Hf) (|1.408| eÅ) is also |0.022| eÅ larger than that of the UiO-

66(Zr) (|1.386| eÅ). The total polarization changes |$\Delta P$| of the C (|0.041| eÅ) and H (|0.00052| eÅ) atoms in the UiO-66(Hf) are found to be also slightly larger than those of C (|0.040| eÅ) and H (|0.00051| eÅ) in UiO-66(Zr).

*Discussion*

According to our DART-PFM experiments, most assuredly, both UiO-66(Hf) and UiO-66(Zr) MOFs possess remarkable piezoresponse, while MOFs with Hf as the metal nodes have a generally larger piezoresponse than those with Zr. The existence of butterfly-shaped amplitude loops and the corresponding phase loops in the PFS experiments confirm the piezoresponse and the rotation of the charged dipole, respectively.[50, 51] It has been reported that the calculated PR loops are in good agreement with macroscopic P-E tests.[52] The calculated PR loops demonstrate that UiO-66-type MOFs evidently possess ferroelectric hysteresis switching behavior, while the ferroelectric response of UiO-66(Hf)-type MOF is stronger than that of the UiO-66(Zr)-type MOF nanocrystals. These PR loops observations further conclude that the UiO-66 (Hf/Zr) nanocrystals should not have the *Fm-3m* (225) structures according to the DART-PFM and PFS results because such a highly centrosymmetric crystal structure has no piezo/ferro-electricity literally. Considering the PR loops in PFS experiments probed in the z direction (same direction as applied electric field), the polarization changes in the z direction are mainly concerned in the first-principles calculations. Comparing all the four different elements, it can be seen that the transition metal ion Hf/Zr and O hold the major part of the polarization changes in both of the UiO-66(Hf/Zr). The total polarization change of UiO-66(Hf) is larger than that of its counterpart UiO-66(Zr). Thereby, these findings indicate that the bond polarity tends to affect the ferroelectric polarizations. This is in excellent agreement with the fact that non-covalent bonds are generally polar because of the asymmetric electron density.[46] For example, the polarity of Hf-O bond is found to be much larger than that of the Zr-O bond by the studies on NU-1000(Hf) and NUS-6.[40, 53] Therefore, it is reasonable to propose that UiO-66(Hf) has stronger ferroelectric polarization behaviour than that of the UiO-66(Zr) due to the larger local polarity of the Hf-O bond in the frameworks.

In summary, we have experimentally probed the nanoscale piezo/ferro-electric behaviours and theoretically explained the underlying polarization mechanism residing in UiO-66-type MOFs. For the first time, it is found that both the UiO-66(Hf)-type and UiO-66(Zr)-type nanocrystals demonstrate certain piezoelectricity. These findings evidently pinpoint the likely low symmetric lattice *(F-43m, 216)* in real UiO-66 MOF instead of the highly centrosymmetric *Fm-3m (225)*. Besides, UiO-66 (Hf)-type MOFs show stronger ferroelectricity than that of the UiO-66 (Zr)-type nanocrystals. To unravel the underlying mechanism, first-principles calculations have been performed on the UiO-66 MOFs. The calculated Born effective charges as well as the polarization changes of the atoms in the UiO-66(Hf) are larger than those in the UiO-66(Zr), which are in good agreement with the DART-PFM and PFS experimental results. Based on the systematic exploration of the piezo/ferro-electric properties of UiO-66-type MOFs, this work has disclosed the crystalline structure and underpin the essentials of the polar coordination bonds in designing novel ferroelectric MOFs. Moreover, it is suggested that the polarization of MOFs can be improved by judicious selection of transitional metal ions to form more polar bonds and localized electronic structures in their low symmetric crystals.

*Methods*

**Sample preparation**

The synthesis process of UiO-66-type nanocrystals can be found in Supplementary Note 1 and 2. Around 10 mg UiO-66-type MOF powder was dispersed into 20 mL ethanol.[54]

After mild sonication, 100 μL of the resultant colloidal solution was pipetted onto clean silicon wafer substrate with Platinum (Pt) coating.[54] PFM and PFS measurements were conducted after the substrate was dried out at 80°C for 2 hours.

**DART-PFM measurements**

PFM is one of the most widely used scanning probe microscopy (SPM) method for characterizing the piezoelectric property of materials.[55] PFM is capable of monitoring the surface displacement induced by electric bias of piezoelectric materials. During the measurements, an ac voltage is applied to the conductive tip which contacts with the sample surface. The bias induced deflection of cantilever is detected by a laser beam on a four-quadrant photodiode.

DART-PFM technique uses a feedback loop to adjust the drive frequency of the cantilever to match the contact resonance frequency. Two frequencies across the resonant frequency with the same amplitudes are set as monitoring targets. As shown in Supplementary Fig. 4b, the resulting amplitudes ($A_1$ and $A_2$) are no longer the same when the resonant frequency shifts. The amplitude $A_1$ shifts downward to $A_1'$ and $A_2$ moves to $A_2'$. The feedback loop is responded by changing the drive frequency until the shift of $A_2$-$A_1$ signal is zero again.[56, 57]

In this work, DART-PFM measurements based on commercial scanning probe microscopy system (MFP3D-SA, Asylum Research, USA) are systematically conducted to unravel the piezoelectricity of UiO-66-type MOF nanocrystals under ambient air condition (50% ~ 60% relative humidity and room temperature).

**PFS measurements**

PFS is the technique to acquire the local PR loop from the surface of ferroelectric materials. In PFS, the conductive tip approaches the sample surface vertically (z direction). A deflection set point (trigger force) is used as the feedback. When the set point is reached, both amplitude and phase response loops are acquired by swept the bias.[40]

The switching bias applied to the conductive tip comprises a dc part ($V_{dc}$(t)) and an ac part ($V_{ac}$) (Supplementary Fig. 5). The $V_{ac}$ is the same as the PFM bias. The $V_{dc}$ (t) is comprised of a sequence of pulses with time length of $\tau_1$ (high state or dc-on) separated by intervals of zero bias with time length of $\tau_2$ (low state or dc-off). The waveform has an envelope specified by a triangular wave with maximum amplitude $V_{max}$ and time periodic T.

In this work, PFS measurements are conducted on the scanning probe microscopy system (MFP3D-SA, Asylum Research, USA) to unravel the ferroelectricity of UiO-66-type MOF nanocrystals under ambient air condition (50% ~ 60% relative humidity and room temperature).

**Born effective charges and polarization changes by first-principles calculations**

The models of UiO-66(Hf/Zr) were established using software Materials Studio[58] 2016 (Accelrys, San Diego, CA). First-principles calculations implemented in the Vienna *ab-initio* simulation package (VASP)[59, 60] was used to fully relax the UiO-66(Hf/Zr) for both lattice and ion positions. Generalized gradient approximation (GGA) with the Perdew–Burke–Ernzerhof (PBE) functional[61] was used to describe the exchange-correlation interaction. The core electrons were described by the Projector-augmented wave (PAW) technology.[62] A plane-wave basis kinetic energy cutoff was 500 eV. The convergence criterion of 10$^{-5}$ eV and 10$^{-8}$ eV were used in the structural relaxtion and properties calculations.

The polarization change $\Delta P$ is theoretically defined as[42]

$$\Delta P = P(\tau) - P(0) = \sum_i \frac{e}{\Omega} Z^*(i,u) \Delta u,$$

where $\Delta u$, 0, $\tau$, $i$, and $\Omega$ correspond to the polarization distance, starting structure, ending structure, atom index and volume of unit cell respectively.

**Data availability**

The authors declare that all data supporting the findings of this study are available in the article and in Supplementary Information (SI) file. Additional information is available from the corresponding author upon request.

*Author information*

**Corresponding Author**

*Email: mpezk@nus.edu.sg
Email: chengy@ihpc.a-star.edu.sg

**Notes**

‖These authors contribute equally to this work.

*Acknowledgement*

This work is supported by Ministry of Education (Singapore) through National University of Singapore under the Academic Research Grant (AcRF) R-265-000-495-112. The authors acknowledge the material supplied by Dr. Dan Zhao and Dr. Zhigang Hu from Department of Chemical and Biomolecular Engineering, National University of Singapore, for PFM and PFS measurement. One of the authors (Y.S.) also thanks the postgraduate scholarship from National University of Singapore.

*References*

1. Vatansever, D., Siores, E. & Shah, T. Alternative resources for renewable energy: piezoelectric and photovoltaic smart structures. In: *Global Warming - Impacts and Future Perspective* (InTech, UK, 2012).

2. Lines, M. E. & Glass, A. M. *Principles and Applications of Ferroelectrics and Related Materials* (Oxford Univ. Press, Oxford, 1977).

3. Lous, R. S. Ferroelectric memory devices, how to store information of future. *Top Master Programme in Nanoscience* **23pp**, (2011).

4. Setter, N. *et al.* Ferroelectric thin films: review of materials, properties, and applications. *J. Appl. Phys.* **100**, 051606 (2006).

5. Wersing, W. & Bruchhaus, R. Ferroelectric thin films and their applications. In: *[Proceedings] Second International Conference on Thin Film Physics and Applications* (1994).

6. Malinovsky, V. K. The prospects of ferroelectrics applications. In: *[Proceedings] 1990 IEEE 7th International Symposium on Applications of Ferroelectrics* (1990).

7. Ortega, N., Ashok, K., Scott, J. F. & Ram, S. K. Multifunctional magnetoelectric materials for device applications. *J. Phys. Condens. Matter.* **27**, 504002 (2015).

8. Martin, L. W. & Rappe, A. M. Thin-film ferroelectric materials and their applications. *Nat. Rev. Mater.* **2**, 16087 (2016).

9. Farha, O. K. *et al.* Metal–organic framework materials with ultrahigh surface areas: is the sky the limit? *J. Am. Chem. Soc.* **134**, 15016-15021 (2012).

10. Deng, H. *et al.* Large-pore apertures in a series of metal-organic frameworks. *Science* **336**, 1018-1023 (2012).

11. Furukawa, H. *et al.* Isoreticular expansion of metal–organic frameworks with triangular and square building units and the lowest calculated density for porous crystals. *Inorg. Chem.* **50**, 9147-9152 (2011).

12. Barea, E., Montoro, C. & Navarro, J. A. R. Toxic gas removal - metal-organic frameworks for the capture and degradation of toxic gases and vapours. *Chem. Soc. Rev.* **43**, 5419-5430 (2014).

13. Sumida, K. *et al.* Carbon dioxide capture in metal–organic frameworks. *Chem. Rev.* **112**, 724-781 (2012).

14. Lee, J., Farha, O. K., Roberts, J., Scheidt, K. A., Nguyen, S. T. & Hupp, J. T. Metal-organic framework materials as catalysts. *Chem. Soc. Rev.* **38**, 1450-1459 (2009).

15. Gascon, J., Corma, A., Kapteijn, F. & Llabrés i Xamena, F. X. Metal organic framework catalysis: Quo vadis? *ACS Catal.* **4**, 361-378 (2014).

16. Kreno, L. E., Leong, K., Farha, O. K., Allendorf, M., Van Duyne, R. P. & Hupp, J. T. Metal–organic framework materials as chemical sensors. *Chem. Rev.* **112**, 1105-1125 (2012).

17. Horcajada, P. *et al.* Metal–organic frameworks in biomedicine. *Chem. Rev.* **112**, 1232-1268 (2012).

18. Medishetty, R. & Vittal, J. J. Metal-organic frameworks for photochemical reactions. In: *Metal-Organic Frameworks for Photonics Applications* (Springer, Berlin, Heidelberg, 2014).

19. Tanabe, K. K., Allen, C. A. & Cohen, S. M. Photochemical activation of a metal–organic framework to reveal functionality. *Angew. Chem. Int. Ed.* **49**, 9730-9733 (2010).

20. Gutierrez, M., Cohen, B., Sanchez, F. & Douhal, A. Photochemistry of Zr-based MOFs: ligand-to-cluster charge transfer, energy transfer and excimer formation, what else is there? *Phys. Chem. Chem. Phys.* **18**, 27761-27774 (2016).

21. LIabres i Xamena, F. & Gascon, J. Characterization of MOFs. 2. Long and local range order structural determination of MOFs by combining EXAFS and diffraction techniques. In: *Metal Organic Frameworks as Heterogeneous Catalysts* (The Royal Society of Chemistry, Cambridge, UK, 2013).

22. Cavka, J. H. *et al.* A new Zirconium inorganic building brick forming metal organic frameworks with exceptional stability. *J. Am. Chem. Soc.* **130**, 13850-13851 (2008).

23. Valenzano, L. *et al.* Disclosing the complex structure of UiO-66 metal organic framework: a synergic combination of experiment and theory. *Chem. Mater.* **23**, 1700-1718 (2011).

24. Wu, H. *et al.* Unusual and highly tunable missing-linker defects in Zirconium metal–organic framework UiO-66 and their important effects on gas adsorption. *J. Am. Chem. Soc.* **135**, 10525-10532 (2013).

25. De Vos, A., Hendrickx, K., Van Der Voort, P., Van Speybroeck, V. & Lejaeghere, K. Missing linkers: an alternative pathway to UiO-66 electronic structure engineering. *Chem. Mater.* **29**, 3006-3019 (2017).

26. Yang, L. -M., Ganz, E., Svelle, S. & Tilset, M. Computational exploration of newly synthesized zirconium metal-organic frameworks UiO-66, -67, -68 and analogues. *J. Mater. Chem. C* **2**, 7111-7125 (2014).

27. Bristow, J. K., Tiana, D. & Walsh, A. transferable force field for metal–organic frameworks from first-principles: BTW-FF. *J. Chem. Theory Comput.* **10**, 4644-4652 (2014).

28. Yang, Q., Wiersum, A. D., Llewellyn, P. L., Guillerm, V., Serre, C. & Maurin, G. Functionalizing porous zirconium terephthalate UiO-66(Zr) for natural gas upgrading: a computational exploration. *Chem. Commun.* **47**, 9603-9605 (2011).

29. Zhao, H. -R., Li, D. -P., Ren, X. -M., Song, Y. & Jin, W. -Q. Larger spontaneous polarization ferroelectric inorganic−organic hybrids: [PbI$_3$]$_\infty$ chains directed organic cations aggregation to Kagomé-shaped tubular architecture. *J. Am. Chem. Soc.* **132**, 18-19 (2010).

30. Noro, S. -I. & Kitagawa, S. *The Supramolecular Chemistry of Organic-Inorganic Hybrid Materials* (John Wiley and Sons, Inc., New Jersey, 2010).

31. Chen, B., Shi, J., Zheng, X., Zhou, Y., Zhu, K. & Priya, S. Ferroelectric solar cells based on inorganic-organic hybrid perovskites. *J. Mater. Chem. A* **3**, 7699-7705 (2015).

32. Rodriguez, B. J., Callahan, C., Kalinin, S. V. & Proksch, R. Dual-frequency resonance-tracking atomic force microscopy. *Nanotechnology* **18**, 475504 (2007).

33. Kalinin, S. V., Rodriguez, B. J. & Kholkin, A. L. Piezoresponse force microscopy and spectroscopy. In: *Encyclopedia of Nanotechnology* (Springer, Netherlands, 2012).

34. Ganpule, C. S. *et al.* Imaging three-dimensional polarization in epitaxial polydomain ferroelectric thin films. *J. Appl. Phys.* **91**, 1477-1481 (2002).

35. Roelofs, A., Böttger, U., Waser, R., Schlaphof, F., Trogisch, S. & Eng, L. M. Differentiating 180° and 90° switching of ferroelectric domains with three-dimensional piezoresponse force microscopy. *Appl. Phys. Lett.* **77**, 3444-3446 (2000).

36. Nonnenmann, S. S., Leaffer, O. D., Gallo, E. M., Coster, M. T. & Spanier, J. E. Finite curvature-mediated ferroelectricity. *Nano Lett.* **10**, 542-546 (2010).

37. Rodriguez, B. J., Jesse, S., Alexe, M. & Kalinin, S. V. Spatially resolved mapping of polarization switching behavior in nanoscale ferroelectrics. *Adv. Mater.* **20**, 109-114 (2008).

38. Rodriguez, B. J. *et al.* Vortex polarization states in nanoscale ferroelectric arrays. *Nano Lett.* **9**, 1127-1131 (2009).

39. Gruverman, A. *et al.* Vortex ferroelectric domains. *J. Phys. Condens. Matter* **20**, 342201 (2008).

40. Sun, Y., Hu, Z., Zhao, D. & Zeng, K. Probing nanoscale functionalities of metal-organic framework nanocrystals. *Nanoscale*, (2017).

41. Roy, A., Prasad, R., Auluck, S. & Garg, A. First-principles calculations of Born effective charges and spontaneous polarization of ferroelectric bismuth titanate. *J. Phys. Condens. Matter* **22**, 165902 (2010).

42. Wang, C. -Z., Yu, R. & Krakauer, H. Polarization dependence of Born effective charge and dielectric constant in KNbO3. *Phys. Rev. B Condens. Matter* **54**, 11161-11168 (1996).

43. Roy, A., Mukherjee, S., Gupta, R., Auluck, S., Prasad, R. & Garg, A. Electronic structure, Born effective charges and spontaneous polarization in magnetoelectric gallium ferrite. *J. Phys. Condens. Matter* **23**, 325902 (2011).

44. Ghosh, S., Di Sante, D. & Stroppa, A. Strain tuning of ferroelectric polarization in hybrid organic inorganic Perovskite compounds. *J. Phys. Chem. Lett.* **6**, 4553-4559 (2015).

45. Sun, Y., Zhuo, Z. & Wu, X. Ferroelectricity and magnetism in metal-formate frameworks of [NH4][M(HCOO)3] (M = Sc to Zn): a first-principles study. *RSC Adv.* **6**, 113234-113239 (2016).

46. Bayne, S. & Carlin, M. *Forensic Applications of High Performance Liquid Chromatography* (CRC Press, USA, 2010).

47. Arlt, G. & Neumann, H. Internal bias in ferroelectric ceramics: origin and time dependence. *Ferroelectrics* **87**, 109-120 (1988).

48. Tayi, A. S., Kaeser, A., Matsumoto, M., Aida, T. & Stupp, S. I. Supramolecular ferroelectrics. *Nat. chem.* **7**, 281-294 (2015).

49. Gonze, X. & Lee, C. Dynamical matrices, Born effective charges, dielectric permittivity tensors, and interatomic force constants from density-functional perturbation theory. *Phys. Rev. B* **55**, 10355-10368 (1997).

50. Hu, W. J. *et al.* Universal ferroelectric switching dynamics of vinylidene fluoride-trifluoroethylene copolymer films. *Sci. Rep.* **4**, 4772 (2014).

51. Choi, Y. -Y. *et al.* Enhancement of local piezoresponse in polymer ferroelectrics via nanoscale control of microstructure. *ACS Nano* **9**, 1809-1819 (2015).

52. Rodriguez, B. J., Jesse, S., Seal, K., Balke, N., Kalinin, S. V. & Proksch, R. Dynamic and spectroscopic modes and multivariate data analysis in piezoresponse force microscopy. In: *Scanning Probe Microscopy of Functional Materials* (Springer, New York, USA, 2010).

53. Beyzavi, M. H. *et al.* A hafnium-based metal–organic framework as an efficient and multifunctional catalyst for facile CO2 fixation and regioselective and enantioretentive epoxide activation. *J. Am. Chem. Soc.* **136**, 15861-15864 (2014).

54. Sun, Y., Hu, Z., Zhao, D. & Zeng, K. Mechanical properties of microcrystalline metal-organic frameworks (MOFs) measured by bimodal amplitude modulated-frequency modulated atomic force microscopy. *ACS Appl. Mater. Interfaces* **9**, 32202-32210 (2017).

55. Kholkin, A. L., Kalinin, S. V., Roelofs, A. & Gruverman, A. Review of ferroelectric domain imaging by piezoresponse force microscopy. In: *Scanning Probe Microscopy: Electrical and Electromechanical Phenomena at the Nanoscale* (Springer, New York, USA, 2007).

56. Gannepalli, A., Yablon, D. G., Tsou, A. H. & Proksch, R. Mapping nanoscale elasticity and dissipation using dual frequency contact resonance AFM. *Nanotechnology* **22**, 355705 (2011).

57. Gannepalli, A., Yablon, D. G., Tsou, A. H. & Proksch, R. Corrigendum: mapping nanoscale elasticity and dissipation using dual frequency contact resonance AFM. *Nanotechnology* **24**, 159501 (2013).

58. Meunier, M. *Guest editorial: Materials Studio* (2008).

59. Kresse, G. & Furthmüller, J. Efficient iterative schemes for ab initio total-energy calculations using a plane-wave basis set. *Phys. Rev. B* **54**, 11169-11186 (1996).

60. Kresse, G. & Furthmüller, J. Efficiency of ab-initio total energy calculations for metals and semiconductors using a plane-wave basis set. *Comp. Mater. Sci.* **6**, 15-50 (1996).

61. Perdew, J. P., Burke, K. & Ernzerhof, M. Generalized gradient approximation made simple. *Phys. Rev. Lett.* **78**, 1396-1396 (1997).

62. Blöchl, P. E. Projector augmented-wave method. *Phys. Rev. B* **50**, 17953-17979 (1994).